\documentclass{aa}
\usepackage{amsmath,graphics,amssymb}
\usepackage[varg]{txfonts}
\usepackage{natbib}
\usepackage{times}
\usepackage{booktabs}
\usepackage[hidelinks]{hyperref}

\newfont{\eufont}{eufm10}

\newcommand{\zav}[1]{\left(#1\right)}
\newcommand{\hzav}[1]{\left[#1\right]}

\newcommand{\kms}{\ensuremath{\text{km}\,\text{s}^{-1}}}

\newcommand{\ms}{\ensuremath{{M}_{\odot}}}
\newcommand{\msr}{\ensuremath{\ms\,\text{yr}^{-1}}}
\newcommand\vnek{\ensuremath{\varv_\infty}}

\newcommand{\de}{\mathrm{d}}
\newcommand{\teff}{\ensuremath{T_\text{eff}}}

\newcommand\Gammac{\Gamma_\text{c}}

\newcommand{\hv}{\tablefootmark{$\ast$}}
\newcommand{\hvd}{\tablefootmark{$\ast\ast$}}

\begin{document}

\title{New line-driven mass-loss rate predictions for central stars\\ of planetary
nebulae: Luminosity dependence\\ when approaching the Eddington limit}
\titlerunning{New mass-loss rate predictions for central stars of planetary
nebulae}

\author{J.~Krti\v{c}ka\inst{1} \and J.~Kub\'at\inst{2} \and
        I.~Krti\v ckov\'a\inst{1}}

\institute{Department of Theoretical Physics and Astrophysics, Faculty of
           Science, Masaryk University, CZ-611 37 Brno,
           Czech Republic \and
           Astronomical Institute, Czech Academy of Sciences, CZ-251 65
           Ond\v rejov, Czech Republic }

\date{Received}

\abstract{After expelling most of their envelopes, asymptotic giant branch stars
turn bluewards in the Hertzsprung-Russel diagram and become hot objects. These
objects ionise the surrounding material, possibly creating a planetary nebula.
Because they are hot and rather luminous, they are able to launch a line-driven wind. The
luminosity dependence of these winds has not yet been studied in detail using
global wind models that consistently consider the interaction of the photosphere and
the wind. We provide line-driven wind models of the central stars of planetary nebulae
focusing on the dependence of these winds on luminosity. While the mass-loss
rate is strongly sensitive to stellar luminosity, the terminal velocity increases
with effective temperature and only weakly depends on luminosity. The predicted
values agree reasonably well with observations, except that observational
mass-loss rates show a rather large scatter, and the predicted terminal velocities are
slightly lower than observational values. We provide ionising fluxes and show
that for stars cooler than approximately 40\,kK, the ionising fluxes computed
from plane-parallel models are inadequate. The wind efficiency stays rather weak, which
implies that the Wolf-Rayet type of central stars is connected with objects with an
elevated mass fraction of heavy elements.}


\maketitle
\nolinenumbers

\begin{table*}
\caption{Comparison of the wind parameters calculated with a restricted
continuum radiative force with models without restriction for stars
corresponding to the evolutionary track of \citet{vasiwo} with a main-sequence
mass of $1\,M_\odot$ and a CSPN mass of $0.569\,M_\odot$}.
\centering
\label{hvezparlim}
\begin{tabular}{lrrccccrcr}
\hline
\hline
Model&\multicolumn{5}{c}{Stellar parameters} & \multicolumn{2}{c}{Restricted} &
\multicolumn{2}{c}{New models}\\
\cmidrule(lr){2-6} \cmidrule(lr){7-8} \cmidrule(lr){9-10}
&\multicolumn{1}{c}{$\teff$} & \multicolumn{1}{c}{$R_{*}$} & $\log g$ &
$\log(L/L_\odot)$ & $\Gamma_\text{e}$ & $\dot M$ & \vnek & $\dot M$ & \vnek \\
& \multicolumn{1}{c}{$[\text{K}]$} & \multicolumn{1}{c}{$[{R}_{\odot}]$} &
cgs & & & [\msr] & [\kms] & [\msr] & [\kms] \\
\hline
M1T10\hv &  10000 & 19.6882 & 1.60 & 3.541 & 0.14 & $1.7\times10^{-10}$& 130 & $3.2\times10^{-10}$&  60 \\
M1T12    &  12388 & 12.8738 & 1.97 & 3.544 & 0.14 &                    &     & $4.0\times10^{-10}$& 160 \\
M1T15    &  15311 &  8.4665 & 2.34 & 3.548 & 0.15 & $1.0\times10^{-8}$ &  90 & $8.5\times10^{-9}$ &  60 \\
M1T18    &  18967 &  5.5425 & 2.71 & 3.552 & 0.15 & $2.0\times10^{-8}$ & 230 & $1.7\times10^{-8}$ & 110\\
M1T23    &  23442 &  3.6451 & 3.07 & 3.556 & 0.15 & $3.8\times10^{-9}$ & 270 & $3.8\times10^{-9}$ & 270\\
M1T29    &  29040 &  2.3834 & 3.44 & 3.559 & 0.16 & $2.4\times10^{-9}$ & 540 & $2.3\times10^{-9}$ & 550\\
M1T35    &  35975 &  1.5513 & 3.81 & 3.558 & 0.16 & $3.8\times10^{-9}$ & 590 & $3.8\times10^{-9}$ & 620\\
M1T44    &  44463 &  1.0097 & 4.19 & 3.553 & 0.17 & $6.0\times10^{-9}$ & 760 & $6.5\times10^{-9}$ & 760\\
M1T55\hvd&  55081 &  0.6497 & 4.57 & 3.542 & 0.16 &                    &     & $4.7\times10^{-9}$ &1150\\
M1T67    &  67920 &  0.4166 & 4.95 & 3.520 & 0.15 & $3.0\times10^{-9}$ &1550 & $3.2\times10^{-9}$ &1520\\
M1T83    &  83176 &  0.2641 & 5.35 & 3.476 & 0.14 & $2.9\times10^{-9}$ &2630 & $3.9\times10^{-9}$ &2450\\
M1T99\hvd&  99083 &  0.1695 & 5.74 & 3.395 & 0.12 &                    &     & $1.3\times10^{-9}$ &2530\\
M1T111   & 111429 &  0.1170 & 6.06 & 3.277 & 0.09 & $1.2\times10^{-10}$&3870 & $1.0\times10^{-9}$ &2610\\
M1T116   & 116413 &  0.0915 & 6.27 & 3.139 & 0.06 & $6.1\times10^{-10}$&2330 & $6.1\times10^{-10}$&2350\\
M1T117\hvd&117761 &  0.0761 & 6.43 & 2.999 & 0.05 &                    &     & $2.6\times10^{-10}$&2170\\
\hline
\end{tabular}
\tablefoot{\tablefoottext{$\ast$}{Model with $\tau_\text{Ross}(R_*)=2/3$.}
\tablefoottext{$\ast\ast$}{Model from \citet{btvit}, which already
accounts for an unrestricted continuum radiative force.}}
\end{table*}

\section{Introduction}

Stellar winds driven by light absorption in lines of heavy elements are
typically associated with hot, luminous, massive stars \citep{lusol,cak}.
However, even low-mass stars, which spend most of their life as cool objects,
can have line-driven winds during certain phases of their evolution. Hot
subdwarfs and central stars of planetary nebulae (CSPNe) are typical examples of
such evolutionary phases, during which low-mass stars can become hot and
luminous enough to drive line-driven outflows \citep{btpaupuv,prisam,vinca,un}.

After exhausting their available helium fuel at the end of the asymptotic giant
branch (AGB) phase, low-mass star remnants are left with just a tiny hydrogen
envelope. The hydrogen burning in this layer maintains the high luminosity of the
remnant \citep{prehlasci,vasiwo} and drives the evolution towards extremely high
temperatures as nuclear active layers burn throughout the leftover hydrogen
towards the stellar surface. The hot central star ionises the outflows left
during the AGB phase \citep{pekny2}. If the expansion time of the envelope is
comparable to the transition timescale, a magnificent planetary nebula is
created \citep{kwoksan}. This evolutionary phase quickly ends, extinguishing the
hydrogen burning and causing the star to settle onto the white dwarf cooling
track in the Hertzsprung-Russell diagram. 

The CSPN phase is rather brief, with a duration of hundreds to thousands of
years \citep{vasiwo,milbe}. This would imply that planetary nebulae are rather
rare objects. However, they correspond to the final evolutionary phases of
solar-type stars, which are one of the most common objects in stellar
populations \citep{mnam}. As a result, thousands of planetary nebulae have been
observed in our Galaxy and in nearby galaxies
\citep{perko,menteok,andromedaplan,fornax}.

The traditional perception of CSPNe as products of single-star evolution
transitioning from the AGB phase to the realm of white dwarfs \citep{nesvoboda}
was challenged by the detection of a large fraction of binary CSPNe
\citep{alltess,chor,nestahnu}. Moreover, the binary scenario is frequently
invoked to explain bipolar morphologies often observed among planetary nebulae
\citep{boffint,kwittova}. However, in any case, hot central objects reside in
the centres of planetary nebulae. The winds of these objects can be strong
enough to shape the nebula and affect the post-AGB timescales of
higher-metallicity central stars \citep{milbe}.

Stellar winds of hot, massive stars can be probed observationally across nearly
the complete electromagnetic spectrum, from X-rays to radio. Probes of the stellar
winds of CSPNe are not so diverse, but still include X-ray emission
\citep{chu,guesbor} and P~Cygni line spectroscopy in the ultraviolet and optical
domains \citep{kuplan,herbian,kasi}. The main aim of the stellar wind studies is
to understand the effect of mass loss on the stellar evolution and the
surrounding environment. To a large extent, this can be described by just two
wind parameters, that is, the mass-loss rate, and the terminal velocity. The mass-loss
rate gives the amount of mass that is returned from the star to the circumstellar
environment per unit of time. On the other hand, the wind terminal velocity
gives the wind speed at large distances from the star. Given the importance of
these parameters, a significant fraction of the stellar wind research focuses on
their determination.

While the observational determination of the terminal velocity from the
absorption edge of P~Cygni line profiles is straightforward \citep{prina,lsl}
and becomes problematic only in certain circumstances with a weak effect on the
final value \citep{hezkysedi}, the observational determination of the mass-loss
rate is much more intricate. The main problem is that at high densities, the
absorption parts of the P~Cygni line profiles of the dominant ions become
saturated, meaning that the analysis must rely on lines originating from minor
ionisation states.  Therefore, this approach requires knowledge of the
ionisation state of the matter, which needs to be obtained using dedicated
spectral modelling codes, such as \mbox{CMFGEN} \citep{hilmi}, FASTWIND
\citep{fastpuls}, and PoWR \citep{powr}. Furthermore, the ionisation state and
radiative transfer are sensitive to the small-scale wind structure, known as
clumping. This has to be taken into account during the spectral analysis
\citep{chuchcar,clres1}. In this respect, great strides have been made via the
ULLYSES \citep{ulisne2} and XShootU \citep{xshootu} projects, which provide
the homogeneous datasets required for a precise wind analysis.

Although hydrodynamic simulations indicate that the wind structure does not
significantly affect the mean mass-loss rate \citep{deb2d}, there are other
subtleties in theoretical models. Precise wind mass-loss rate predictions have
to account for a detailed radiative transfer from the stellar photosphere to the
stellar wind. This procedure requires global (unified) wind models
\citep{grahamz,cmfkont,powrdyn,sundyn}.

The application of wind spectrum synthesis codes that account for clumping has
led to a significant downward revision of the wind mass-loss rates of OB stars,
by a factor of about three \citep[e.g.][]{hezkysedi,hawvmm}. A similar
reduction is also predicted by global models, bringing theory and observation
into reasonable agreement \citep{zcmfkont}. However, the application of global
models in the CSPN domain is currently limited to a single evolutionary track
\citep{btvit}. This is a significant shortcoming, especially because the luminosities of some
more massive central stars can be so high that they approach
the Eddington limit due to continuum opacity. To address this issue, we
supplemented the mass-loss rate predictions of \citet{btvit} by providing wind
models for CSPNs with high luminosities.

\begin{table*}
\caption{Predicted wind parameters for CSPNe with high luminosities and
main-sequence masses $1.25\,M_\odot$ (CSPN mass $0.562\,M_\odot$) and
$2\,M_\odot$ (CSPN mass $0.58\,M_\odot$).}
\centering
\label{hvezpar}
\begin{tabular}{lrrccccr}
\hline
\hline
Model &\multicolumn{1}{c}{$\teff$} & \multicolumn{1}{c}{$R_{*}$} & $\log g$ &
$\log(L/L_\odot)$ & $\Gamma_\text{e}$ & $\dot M$ & \vnek \\
& \multicolumn{1}{c}{$[\text{K}]$} & \multicolumn{1}{c}{$[{R}_{\odot}]$} &
cgs & & & [\msr] & [\kms] \\
\hline
M1.25T10\hv&10057 & 24.14  & 1.42 & 3.728 & 0.21 & $1.2\times10^{-9}$ & 40 \\
M1.25T15&   15494 & 10.15  & 2.18 & 3.726 & 0.21 & $2.3\times10^{-9}$ & 30 \\
M1.25T18&   18719 &  6.943 & 2.50 & 3.725 & 0.23 & $2.6\times10^{-8}$ &  130 \\
M1.25T23&   23233 &  4.494 & 2.88 & 3.722 & 0.23 & $3.4\times10^{-8}$ &  270 \\
M1.25T28&   28837 &  2.905 & 3.26 & 3.719 & 0.23 & $5.1\times10^{-9}$ &  100 \\
M1.25T35&   35792 &  1.875 & 3.64 & 3.714 & 0.23 & $1.0\times10^{-8}$ &  380 \\
M1.25T44&   44425 &  1.208 & 4.02 & 3.707 & 0.24 & $1.0\times10^{-8}$ &  710 \\
M1.25T68&   68439 &  0.493 & 4.80 & 3.679 & 0.23 & $6.1\times10^{-9}$ & 1170 \\
M1.25T99&   99889 &  0.215 & 5.52 & 3.616 & 0.19 & $2.9\times10^{-9}$ & 2500 \\
M1.25T111& 111576 &  0.164 & 5.76 & 3.573 & 0.18 & $3.1\times10^{-9}$ & 2600 \\
\hline
M2T10\hv& 10065 & 28.3   & 1.30 & 3.867 & 0.29 & $4.5\times10^{-9}$ & 30 \\
M2T15\hv& 15175 & 12.4   & 2.01 & 3.864 & 0.30 & $6.7\times10^{-8}$ & 120 \\
M2T18\hv& 18633 &  8.24  & 2.37 & 3.866 & 0.31 & $1.1\times10^{-7}$ & 80 \\
M2T23  &  23559 &  5.13  & 2.78 & 3.861 & 0.31 & $9.4\times10^{-8}$ & 200 \\
M2T28  &  28927 &  3.39  & 3.14 & 3.858 & 0.30 & $1.9\times10^{-8}$ & 290 \\
M2T35  &  35519 &  2.242 & 3.50 & 3.856 & 0.30 & $1.7\times10^{-8}$ & 180 \\
M2T44  &  44910 &  1.394 & 3.91 & 3.850 & 0.32 & $2.0\times10^{-8}$ & 510 \\
M2T67  &  67708 &  0.603 & 4.64 & 3.836 & 0.31 & $1.2\times10^{-8}$ & 930 \\
M2T99  &  99129 &  0.272 & 5.33 & 3.807 & 0.29 & $6.7\times10^{-9}$ & 2470 \\
M2T111 & 111466 &  0.210 & 5.56 & 3.786 & 0.28 & $6.1\times10^{-9}$ & 2570 \\
\hline
\end{tabular}
\tablefoot{\tablefoottext{$\ast$}{Model with $\tau_\text{Ross}(R_*)=2/3$.}}

\end{table*}

\section{Wind models}

The winds of CSPNe are mostly accelerated by the absorption of stellar radiation
in the lines of elements heavier than helium. Therefore, a detailed solution of
the radiative transfer equation is needed for all relevant transitions, from the
ultraviolet to the infrared regions, in order to derive precise wind models.
Moreover, part of the stellar flux is absorbed in the wind and is re-emitted
towards the photosphere, where it can affect the temperature and density
distribution. In consequence, precise models have to describe the
photosphere and the stellar wind in the same unified manner. Several such models
are available in the literature \citep{grahamz,powrdyn,sundyn}.

We applied our METUJE global spherically symmetric wind code to model the stellar
winds of CSPNe \citep{cmfkont}. The code assumes a spherically symmetric,
stationary, and smooth flow and solves the hydrodynamic, radiative transfer,
and kinetic (statistical) equilibrium equations to derive the radial variations
of density, velocity, temperature, level populations, and specific intensities.
The radiative force, which includes contributions from the continuum and lines,
and the radiative heating and cooling term \citep{kpp} are calculated using the
solution of the radiative transfer equation in the comoving frame \citep{mikuh}.
The resulting radiation field, together with collisional terms, also determines
the level populations, which are derived from kinetic equilibrium equations
\citep [also known as NLTE equations;][]{hubenymihalas}. These equations are
solved for the elements and ionisation states listed in \citet{btvit}. The
corresponding ionic models are either taken from TLUSTY atmosphere models
\citep{ostar2003,bstar2006} or prepared by us using data from the Opacity and
Iron Projects \citep{topt,zel0} and corrected using observational data derived
from the NIST database \citep{nist}. 

The model equations were supplemented by the relevant boundary conditions. While
these conditions are usually straightforward, the definition of the stellar
radius is more problematic. In typical model atmospheres, the stellar radius is
defined by the internal boundary condition, requiring the optical depth to reach
a prescribed value at some internal model point. To avoid numerical problems,
most of our models calculated so far have defined the stellar radius as the
radius of the lowermost model point. While the difference between these two
implementations is small for high-gravity geometrically thin atmospheres, this
approach can be problematic in extended atmospheres close to the Eddington
limit. To address this, we introduced an internal boundary condition for the
stellar radius $R_\ast$ in the form of
\begin{equation}
\label{slozitarovnice}
\tau_\text {Ross} (R_\ast)=\frac{2}{3},
\end{equation}
where $\tau_\text {Ross} $ is the Rosseland optical depth \citep{hubenymihalas}.
Our numerical tests showed that the application of Eq.~\eqref{slozitarovnice}
does not significantly modify the wind solution in geometrically thin
atmospheres. In contrast to our expectations, Eq.~\eqref{slozitarovnice} even leads
to a better convergence of the models, perhaps due to smaller variations of
photospheric parameters during individual iterations. We used
Eq.~\eqref{slozitarovnice} only for the most extended models, which are denoted
by an asterisk in Tables~\ref{hvezparlim} and \ref{hvezpar}.

METUJE solves the model equations iteratively. The initial estimate of the NLTE
photospheric structure and the reference emergent flux were derived from the
TLUSTY code \citep{ostar2003,bstar2006}. The models were parametrised in terms
of the stellar effective temperature \teff, mass $M$, radius $R_*$, and chemical
composition. Except for metallicity, these parameters were taken from post-AGB
evolutionary tracks. As we focused on hydrogen-rich central stars, we assumed a
solar chemical composition after \citet{asp09}. In addition to a consistent wind model,
we derived the wind mass-loss rate $\dot M$, terminal velocity $v_\infty$, and
emergent flux from the final solution.

\section{Low-luminosity models with unlimited continuum radiative force}

In photospheric layers, the continuum radiative force due to bound-free,
free-free, and scattering opacities can be comparable to the gravitational
force. This leads to the inflation of the photosphere, which can cause numerical
problems. To address this issue and improve the convergence of the models,
\citet{btvit} limited the continuum radiative force to twice the radiative force
due to light scattering on free electrons. However, it is unclear to which extent
this restriction affects the wind parameters.

\begin{figure}
\includegraphics[width=0.5\textwidth]{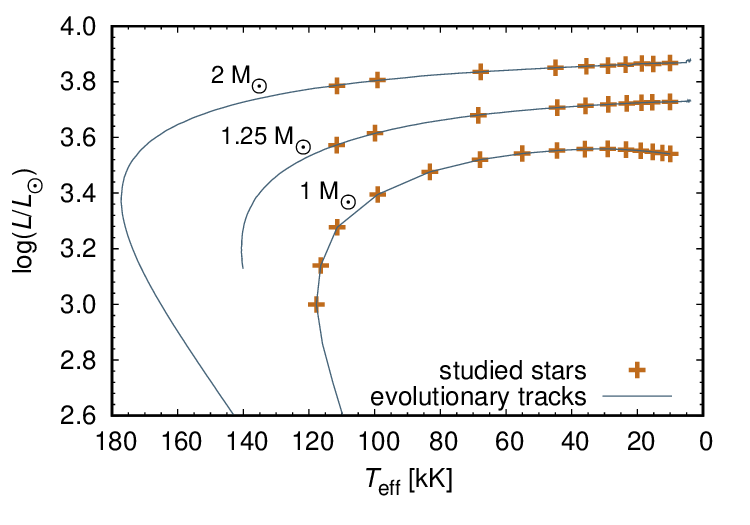}
\caption{Parameters of the model stars in the diagram relating stellar
luminosity to the effective temperature, overplotted with the evolutionary tracks
of \citet{vasiwo} and \citet{milbe}. The tracks are labelled with the initial
stellar masses.}
\label{chlazlbtvit}
\end{figure}

To understand the effect of the continuum radiative force on the wind models, we
recalculated the wind models of \citet{btvit}, lifting the restrictions imposed
on the radiative force. The adopted stellar parameters (given in
Table~\ref{hvezparlim}) were taken from the post-AGB evolutionary tracks of
\citet{vasiwo} with a main-sequence mass of $1\,M_\odot$ and a CSPN mass of
$M=0.569\,M_\odot$. These parameters cover the post-AGB evolutionary tracks up
to the turning point at which the star settles on the white dwarf cooling track
(as illustrated in Fig.~\ref{chlazlbtvit}).

Table~\ref{hvezparlim} compares the wind parameters calculated using the
restricted continuum radiative force with new models calculated without this
restriction. Apparently, neither the wind mass-loss rate nor the terminal velocity
changes significantly when a more realistic atmospheric structure is taken into
account. The wind mass-loss rate depends primarily on stellar luminosity because
a higher luminosity leads to stronger radiative acceleration \citep{cak,pusle}.
This is unaffected by envelope inflation. The inflation modifies
the photospheric temperature only slightly, which in turn modifies the wind mass-loss rate. As
the inflation is weak and does not significantly modify the radius, the terminal
velocity, which mostly scales with the escape speed \citep{cak,lsl}, also
remains nearly unchanged.

\section{CSPNe with the full range of luminosities}

To understand the effect of the proximity to the Eddington limit, we
calculated additional wind models for stars with higher luminosities. In the
CSPNe domain, a higher luminosity implies a higher initial (main-sequence) mass.
For this study, we selected evolutionary tracks with initial masses of
$1.25\,M_\odot$ and $2\,M_\odot$ from the grid of models calculated by
\citet[Fig.~\ref{chlazlbtvit}]{milbe}. These initial parameters give CSPN masses
of $0.562\,M_\odot$ and $0.580\,M_\odot$, respectively.

The parameters of the high-luminosity model stars we studied are given in
Table~\ref{hvezpar}. In addition to the basic stellar parameters, we included
the Eddington parameter $\Gamma_\text{e}$, which gives the ratio of the
radiative force due to free-electron scattering and the gravitational force,
\begin{equation}
\Gamma_\text{e}=\frac{\sigma_\text{e}}{4\pi c G}
\frac{n_\text{e}}{\rho}\frac{L}{M},
\end{equation}
where $\sigma_\text{e}$ is the Thomson scattering cross-section, $n_\text{e}$ is
the electron number density, $\rho$ is the mass density, and $L$ is the stellar
luminosity. Because the ionisation state varies within the stellar envelope, the
ratio of the electron number density and mass density $n_\text{e}/\rho$ is not
constant. For the relatively high effective temperatures considered here, hydrogen
remains ionised. Consequently, changes in $n_\text{e}/\rho$ are mostly due to
changes in helium ionisation. Tables~\ref{hvezparlim} and \ref{hvezpar} evaluate
the Eddington parameter at the wind critical point.

\begin{figure}
\includegraphics[width=0.5\textwidth]{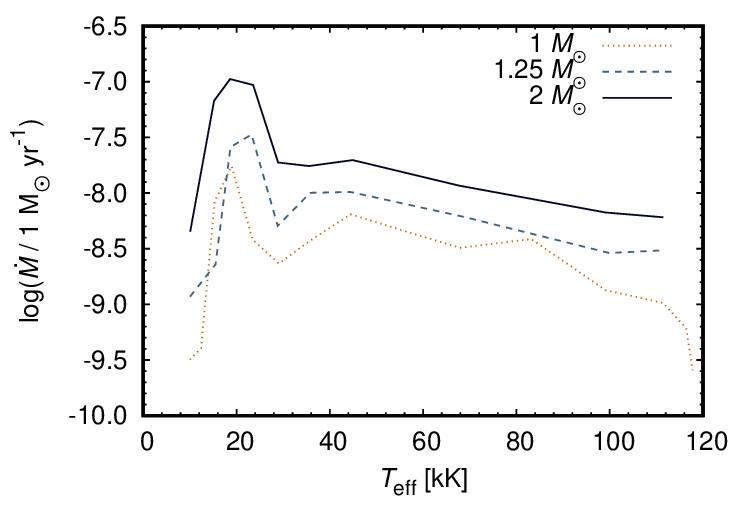}
\caption{Wind mass-loss rate as a function of the effective temperature for
CSPNe with different main-sequence masses.}
\label{lbtdmdttep}
\end{figure}

\begin{table*}[t]
\caption{Fit parameters for the mass-loss rate in Eq.~\eqref{dmdtlbt}.}
\label{parametry}
\centering
\begin{tabular}{*{12}{c}}
\hline
\hline
$a$ & $b$ & $\zeta$ & $Z_\odot$ &
$c_1$ & $c_2$ & $c_3$ & $c_4$ &
$c_{1,l}$ & $c_{2,l}$ & $c_{3,l}$ & $c_{4,l}$ \\
\hline
$-12.060$ &3.95& 0.53 & 0.0134 & 1.73 & 1.67 & 3.04 & 1.35 & $ -0.34$ &
$ -0.98$ & $ -2.69$ & $ -1.29$ \\
\hline
\end{tabular}\medskip

\begin{tabular}{*{8}{c}}
\hline
\hline
$T_1$ & $T_2$ & $T_3$ & $T_4$ &
$\Delta T_1$ & $\Delta T_2$ & $\Delta T_3$ & $\Delta T_4$ \\
\hline
19.44 & 40.0 & 82.1 & 115.6 & 6.63 & 19.2 & 37.3 & 7.96\\
\hline
\end{tabular}
\tablefoot{$T_i$ and $\Delta T_i$ are expressed in units of kK.}
\end{table*}

The mass-loss rate primarily depends on the stellar luminosity (and subsequently,
on the initial mass). Figure~\ref{lbtdmdttep} shows that the mass-loss rate also
depends on the effective temperature, and it therefore varies during the CSPN
evolution. The mass loss first increases as the stars evolve towards higher
effective temperatures because iron becomes increasingly important for radiative
acceleration. After the ionisation of \ion{Fe}{iii} to the less efficient
\ion{Fe}{iv}, the importance of iron to the radiative acceleration slightly
decreases, leading to a decrease in the mass-loss rate. This decrease from the
maximum appears at about $\teff=20\,\text{kK}$ and is termed the bistability jump
\citep{mnichov8,vikolabis,bcmfkont}. A further decrease in the mass-loss rate
towards higher effective temperatures \citep{btpau,btvit} is primarily caused by
the lower efficiency of highly ionised species to accelerate the stellar wind
and partly by the decrease in the luminosity in later evolutionary stages.
However, at even higher effective temperatures than considered here, the
mass-loss rate is expected to rise again with the temperature as a result of
the ionisation of iron to ions higher than \ion{Fe}{ix}. These ions, connected
with so-called iron bump opacities, are able to either directly drive a wind
\citep{nulam,graham,wrtep} or cause strong turbulence \citep{jian,deb2d}.

The mass-loss rate of CSPNe, in the considered range of effective temperatures,
$\teff=10-120\,\text{kK}$, and luminosities, $\log(L/L_\odot)=3.5-3.9$, can be
roughly fitted as
\begin{multline}
\label{dmdtlbt}
\log\zav{\frac{\dot M}{1\, \msr }}= a + b \log\zav{\frac{L}{10^3L_\odot}}+
\zeta\log\zav{\frac{Z}{Z_\odot}}\\
+\sum_{i=1}^4\hzav{c_i+c_{i,l}\log\zav{\frac{L}{10^3L_\odot}}}
\exp\hzav{-\frac{\zav{\teff-T_i}^2}{\Delta T_i^2}},
\end{multline}
where $\log$ denotes the decadic logarithm, and $\exp$ denotes the natural
exponential function. The parameters of the fit are listed in
Table~\ref{parametry}. The metallicity dependence, described via the $\zeta$
parameter, was adopted from \citet{btvit}. The fit \eqref{dmdtlbt} describes the
available predictions reasonably well with a root mean square of the residuals
of about $0.18$.

\begin{figure}
\includegraphics[width=0.5\textwidth]{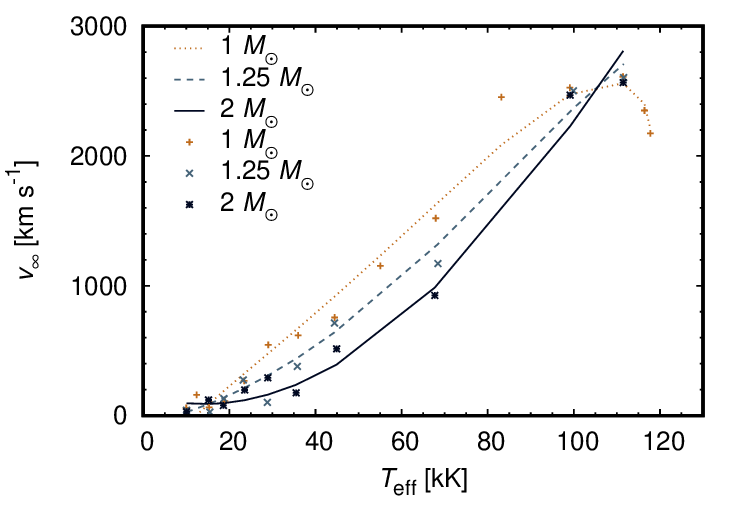}
\caption{Predicted wind terminal velocities as a function of the effective
temperature (points) and their fit via Eq.~\eqref{vnekteplbtfit} (lines).}
\label{vnekteplbt}
\end{figure}

\begin{table}[hbt]
\caption{Parameters
of the terminal velocity fit Eq.~\eqref{vnekteplbtfit}.}
\label{vnekteplbttab}
\centering
\begin{tabular}{cccccccc}
\hline\hline
$v_0$ & $v_{0,l}$ & $v_1$ & $v_{1,l}$ &$v_2$ & $v_{2,l}$ & $\tau$ & $\Delta\tau$\\
\hline
 2036 & $-1039$ & 1231 & 460 & $-1143$ & 1999 & 65 & 55\\
\hline
\end{tabular}
\tablefoot{Parameters $\tau$ and $\Delta\tau$ are expressed in units of kK.}
\end{table}

The wind terminal velocity scales mostly with the escape speed \citep{cak,pusle}. As
a result of the decrease in the radius over time, the wind terminal velocity
increases during a blueward evolutionary motion of CSPNe
\citep[Fig.~\ref{vnekteplbt} and][]{btpau}. The terminal velocity is
additionally slightly lower for higher luminosities because the
escape speed decreases with increasing radius. In total, the wind terminal velocities for
$\teff=10-120\,\text{kK}$ and $\log(L/L_\odot)=3.5-3.9$ can be reasonably well
fitted via
\begin{equation}
\label{vnekteplbtfit}
\zav{\frac{v_\infty}{1\,\kms}}=\sum_{n=0}^2
\hzav{v_n+v_{n,l}\log\zav{\frac{L}{10^3L_\odot}}}
P_n\zav{\frac{\teff-\tau}{\Delta\tau}},
\end{equation}
where $P_n(x)$ are Legendre polynomials of a degree $n$, $P_n(x) =
\frac{\de^n}{\de x^n} (x^2 -1)^n/(2^n n!)$, and the fit parameters are given in
Table~\ref{vnekteplbttab}.

\begin{table}[t]
\caption{Number of ionising photons per unit of surface area and per unit of
time $\log \zav{Q/1\,\text{cm}^{-2}\,\text{s}^{-1}}$ calculated from the
plane-parallel TLUSTY and global spherically symmetric METUJE models.}
\label{pocqlbt}
\begin{tabular}{lcccccc}
\hline\hline
Model & \multicolumn{3}{c}{TLUSTY} & \multicolumn{3}{c}{METUJE}\\
\cmidrule(lr){2-4} \cmidrule(lr){5-7}
& \ion{H}{i} & \ion{He}{i}& \ion{He}{ii}
& \ion{H}{i} & \ion{He}{i}& \ion{He}{ii}\\
\hline
M1T15  & 18.93 & 14.20 &  1.31 & 20.51 & 13.47 & \\
M1T18  & 20.22 & 16.17 &  5.47 & 21.47 & 17.15 & 5.42 \\
M1T23  & 21.79 & 18.55 &  9.26 & 21.65 & 18.24 & 4.67 \\
M1T29  & 23.12 & 20.76 & 12.51 & 21.91 & 19.34 & 8.71 \\
M1T35  & 23.97 & 23.02 & 16.05 & 23.67 & 21.95 & 11.88 \\
M1T44  & 24.53 & 23.91 & 18.93 & 24.43 & 23.70 & 13.42 \\
M1T55  & 24.97 & 24.54 & 20.73 & 24.92 & 24.46 & 21.01 \\
M1T67  & 25.36 & 25.05 & 22.04 & 25.33 & 24.99 & 22.14 \\
M1T83  & 25.71 & 25.48 & 23.22 & 25.70 & 25.47 & 22.86 \\
M1T99  & 25.99 & 25.81 & 24.41 & 26.00 & 25.81 & 23.93 \\
M1T111 & 26.18 & 26.02 & 24.98 & 26.18 & 26.03 & 24.70 \\
M1T116 & 26.25 & 26.10 & 25.12 & 26.25 & 26.10 & 24.93 \\
M1T117 & 26.27 & 26.12 & 25.12 & 26.28 & 26.13 & 24.97 \\
\hline
M1.25T15&  19.08 & 14.41 & 1.31  & 19.64 & 10.83 &       \\
M1.25T18&  20.28 & 16.21 & 5.46  & 21.14 & 13.80 &       \\
M1.25T23&  21.93 & 18.71 & 9.24  & 22.02 & 17.84 & 2.81 \\
M1.25T28&  23.23 & 20.94 & 12.63 & 22.14 & 19.35 & 8.97 \\
M1.25T35&  24.00 & 23.08 & 16.04 & 23.61 & 21.58 & 11.49 \\
M1.25T44&  24.54 & 23.92 & 19.10 & 24.42 & 23.64 & 13.59 \\
M1.25T68&  25.38 & 25.06 & 22.18 & 25.35 & 25.00 & 22.21 \\
M1.25T99&  26.00 & 25.82 & 24.67 & 26.01 & 25.82 & 24.11 \\
M1.25T111& 26.17 & 26.02 & 25.13 & 26.18 & 26.01 & 24.81 \\
\hline
M2T15  & 19.03 & 14.27 & 0.88  & 20.76 & 13.26 &        \\
M2T18  & 20.41 & 16.37 & 5.57  & 21.21 & 13.69 & 0.27 \\
M2T23  & 22.24 & 19.15 & 9.74  & 21.81 & 15.50 & 2.01 \\
M2T28  & 23.35 & 21.30 & 12.95 & 22.30 & 18.81 & 4.61 \\
M2T35  & 24.02 & 23.11 & 16.20 & 23.43 & 21.23 & 12.94 \\
M2T44  & 24.58 & 23.97 & 19.37 & 24.43 & 23.59 & 14.71 \\
M2T67  & 25.36 & 25.04 & 22.24 & 25.29 & 24.91 & 22.11 \\
M2T99  & 25.98 & 25.80 & 24.79 & 26.00 & 25.80 & 24.11 \\
M2T111 & 26.16 & 26.00 & 25.22 & 26.17 & 26.00 & 24.88 \\
\hline
\end{tabular}
\end{table}

The wind modifies the radiative flux that emerges from the stellar surface.
Specifically, far-UV radiation, which ionises planetary nebulae, is sensitive to
the presence of the stellar wind. This can be seen from Table~\ref{pocqlbt},
which shows the number of \ion{H}{i}, \ion{He}{i}, and \ion{He}{ii} ionising
photons emitted per unit of time from a unit area of the stellar surface, 
\begin{equation}
Q=4\pi\int_{\nu_0}^\infty \frac{H_\nu}{h\nu}\,\de\nu.
\end{equation}
Here, $H_\nu$ is the Eddington flux, and $\nu_0$ is the ionisation frequency.
Compared to plane-parallel models, global models modify the emergent flux by
wind blanketing \citep{acko} and by deviations from spherical symmetry
\citep{kuhumi, grkud, scspn}. The stellar winds simulated here are optically
thin for the mean opacities. Therefore, the radiative flux in the spectral
region of the maximum flux originates from the photosphere and is not
significantly affected by the wind. On the other hand, the flux emitted at
frequencies higher than those of the flux maximum can originate in the wind and
can therefore be strongly affected by wind blanketing.

For the hottest stars, the emergent flux maximum is located above \ion{H}{i} and
\ion{He}{i} ionisation frequencies. Although the global models slightly modify
the flux even in the region of the flux maximum, the total flux is the same. The
number of ionising photons from plane-parallel TLUSTY and global spherically
symmetric METUJE also agrees. However, with decreasing effective temperature the
flux maximum shifts towards lower frequencies, becoming lower than the
ionisation frequency. As a result, the number of ionising photons for cooler
stars starts to differ. This first occurs for the \ion{He}{i} ionising flux at
around $\teff\approx50\,\text{kK}$ and subsequently to the \ion{H}{i} ionising
flux at around $\teff\approx40\,\text{kK}$. The ionisation edge of \ion{He}{ii}
always appears at frequencies higher than or close to the flux maxima, so the
number of \ion{He}{ii} ionising photons calculated by TLUSTY (static
plane-parallel models) and METUJE (spherically symmetric global wind models)
never agree. This is partly due to the back-warming of the photosphere caused by
wind blanketing.

\section{Observational tests}

\subsection{Mass-loss rates}

A variety of observational indicators, spanning nearly the entire
electromagnetic spectrum, are used to determine the mass-loss rates of massive
hot stars. These include X-ray line profiles \citep{cohcar}, ultraviolet,
optical, and near-infrared line spectroscopy \citep{najaro,bouhil,clres2}, wind
bow shocks \citep{kobul}, and infrared and radio excess \citep{pulchuch}.
However, in the case of CSPNe, only ultraviolet and optical lines are typically
used \citep{btpau,kuplan,hulpul,herbian}. To obtain precise mass-loss rate
estimates from these lines, the effect of the wind structure (clumping) has to be
accounted for \citep{chuchcar,clres1}. This effect was neglected in the study of
CSPN winds by \citet{btpau}, while the remaining studies accounted for clumping.

\begin{figure}
\includegraphics[width=0.5\textwidth]{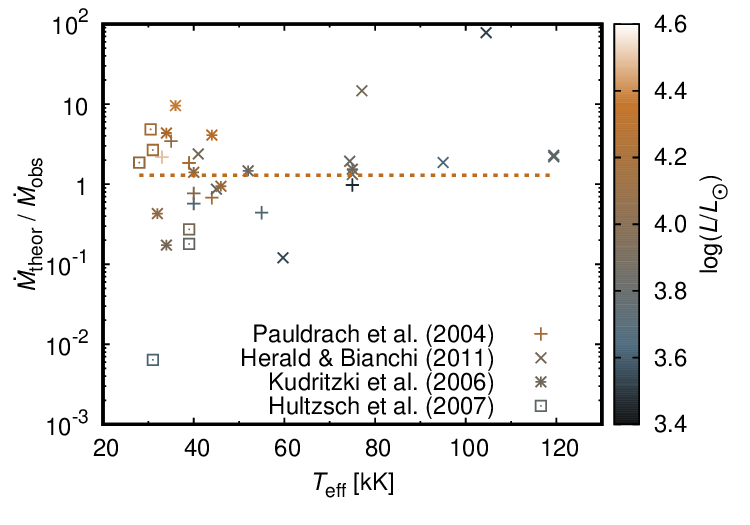}
\caption{Ratio of the wind mass-loss rates of CSPNe determined from predictions
and observations. The plot is coloured according to the stellar luminosity.
The dashed line denotes the mean value.}
\label{dmdtlbt_fig}
\end{figure}

Figure~\ref{dmdtlbt_fig} compares the mass-loss rates predicted by
Eq.~\eqref{dmdtlbt} with observationally determined values. The
predicted and observational values agree reasonably well on average. The average ratio of
predicted and observational mass-loss rates is spoiled by a few offset values,
but the median is around one for all observational datasets. To provide a more
quantitative value, we calculated an exponential from the mean logarithm of the
ratio, which gives a geometric mean ratio of the mass-loss rates of 1.29. This
indicates that the predictions are slightly higher than the
observational values on average. However, the scatter is relatively large, albeit
comparable to that for O stars \citep[e.g.][]{zcmfkont}.

The surface layers of CSPNe can become enriched with heavier elements
synthesised during previous evolutionary stages \citep{milbe}. Because hot star
winds are mainly accelerated by the lines of heavy elements
\citep{cak,abpar,vikola}, the mass-loss rate is sensitive to the metallicity
\citep[e.g.][]{btvit}. The metallicity enhancement is typically accompanied by
a larger abundance of helium, which was determined in the study of \citet{hulpul}.
However, with the exception of one object, this sample shows a nearly solar helium
abundance, implying that self-enrichment is likely to be low. The only
exceptional object, \object{M 1-37}, in the \citet{hulpul} sample that has an
enhanced helium abundance has the highest mass-loss rate despite having the
lowest luminosity. This indicates that the self-enrichment by nuclear reactions
plays a role in the higher mass-loss rate in this case. In addition, the sample
of \citet{hulpul} dealt with bulge objects, which may have very different
metallicites with respect to the solar ones adopted here. 

\subsection{Terminal velocities}

\begin{figure}
\includegraphics[width=0.5\textwidth]{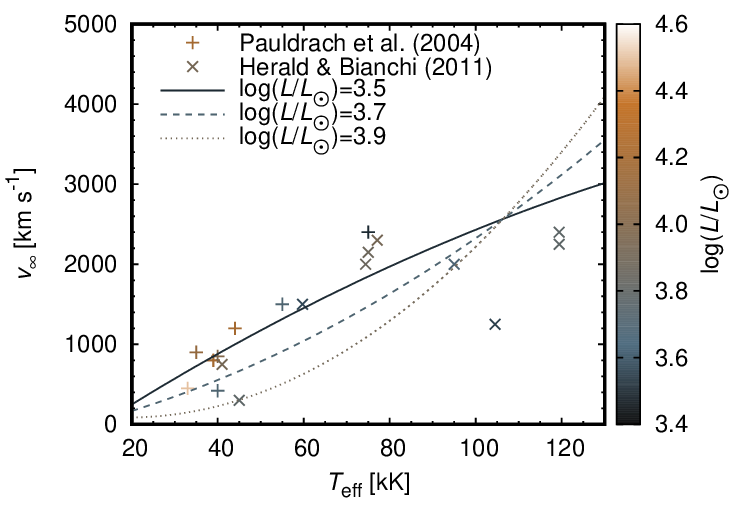}
\caption{Wind terminal velocities predicted for selected values of the CSPN
luminosity from Eq.~\eqref{vnekteplbtfit}, compared to the observational values.
The data points are colour-coded according to the stellar luminosity.}
\label{vnekvlbt}
\end{figure}

In contrast to the mass-loss rates, the wind terminal velocities can be
determined relatively precisely from the blue edge of the absorption component
of the observed P~Cygni line profiles \citep[e.g.][]{hawvmm}. The comparison of the
observationally determined terminal velocities \citep{btpau,herbian} with the
theoretical predictions in Fig.~\ref{vnekvlbt} shows that the theoretical models reproduce the increase in the wind terminal velocity during the
blueward evolutionary motion of CSPNe reasonably well. This increase stems from the
proportionality between the terminal velocity and escape speed \citep{cak} and
from the evolutionary decrease in radius at a nearly constant mass.

Although they reproduce the increasing trend with the effective temperature,
the predictions are typically slightly lower than observational values for
corresponding luminosity. A similar difference emerges in O stars
\citep{zcmfkont} and in the CSPN predictions of \citet{btpau} and \citet{btvit}.
\citet{btpau} suggested that the estimate of the stellar mass should be increased to solve
this discrepancy, while \citet{btvit,zcmfkont} attributed this difference to
model deficiencies, such as the absence of X-rays and radius-dependent clumping
\citep[cf.][]{savi}.

\section{The effect of the continuum radiative force on the photospheric
structure}

\begin{figure}
\includegraphics[width=0.5\textwidth]{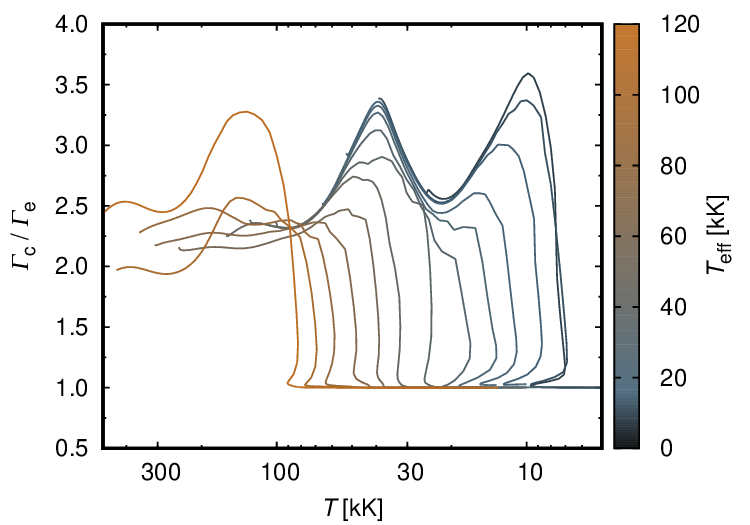}
\caption{Radial variation in the ratio of the total continuum radiative force to
the radiative force due to light scattering on free electrons within the
model atmosphere. We plot this as a function of temperature, which stands as a proxy
for radius, for stars with an initial mass $1\,M_\odot$. The plots are
colour-coded according to the effective temperature. The plots were averaged
over consecutive grid points.}
\label{eddtep}
\end{figure}

Deep within a stellar atmosphere, light scattering on free electrons is not the
only process contributing to the continuum radiative force. Additionally, the
bound-free transitions of the two most abundant elements, hydrogen and helium,
become significant. Therefore, it is reasonable to introduce a generalised
Eddington parameter, $\Gammac$, defined as the ratio of the total continuum
radiative acceleration $g_\text{cont}$ and gravitational acceleration $g$,
\begin{equation}
\label{obedd}
\Gammac =\frac{g_\text{cont}}{g}=\frac{ \overline\kappa F}{cg}=
\frac{\rho\overline\kappa}{\sigma_\text{e}n_\text{e}}\Gamma_\text{e}.
\end{equation}
We have rewritten the generalised Eddington parameter in terms of the
Eddington parameters purely due to electron scattering, the flux mean continuum
opacity per unit of mass $\overline\kappa$, and the flux $F=L/(4\pi R_*^2)$.

The plot showing the radial variation in the ratio of these two Eddington
parameters is given in Fig.~\ref{eddtep} for model stars with initial mass
$1\,M_\odot$. In this plot, temperature acts as a proxy for the location within the
stellar atmosphere, enabling us to compare variations for stars with different
radii. The variations show two bumps at around $10\,\text{kK}$ and
$39\,\text{kK}$, primarily caused by the ionisation of \ion{H}{i} from the $n=2$
and $n=1$ levels, respectively. Plots in deeper regions of the photosphere
accumulate around these bumps. This can be understood from the equation relating
the pressure and temperature variations within the envelope \citep[cf.][]{gredi},
\begin{equation}
\frac{\de p}{\de T}=\frac{16\sigma_\text{R}T^3}{3c}\frac{1-\Gammac}{\Gammac},
\end{equation}
which can be derived from the ratio of the hydrostatic equilibrium equation $\de
p/\de r=\rho g(\Gammac-1)$ and the radiative equilibrium equation in the
diffusion approximation $\de T/\de r=-3cg \rho\Gammac /(16 \sigma_\text{R}T^3)$.
For an ideal gas with $p=\rho k T/(\mu m_\text{H})$ with the help of
Eq.~\eqref{obedd},
\begin{equation}
\label{dpdtrhot}
\frac{k}{\mu m_\text{H}}\frac{\de \zav{\rho T}}{\de T}=
\frac{16\sigma_\text{R}T^3}{3c}
\frac{1-\frac{\rho\overline\kappa}{\sigma_\text{e}n_\text{e}}\Gamma_\text{e}}
{\frac{\rho\overline\kappa}{\sigma_\text{e}n_\text{e}}\Gamma_\text{e}}.
\end{equation}
Because the mean opacity $\overline\kappa$ is a function of $\rho$ and $T$ for a
given chemical composition, Eq.~\eqref{dpdtrhot} shows that deep in the
photosphere, density and pressure become functions only of $T$ and
$\Gamma_\text{e}$. As a result, different model atmospheres with the same
$\Gamma_\text{e}$ show the same dependence of $\rho$ and $p$, and therefore, also
$ \Gammac$ as a function of temperature (Fig.~\ref{eddtep}).

In the upper part of the photospheres, where $\Gamma_\text{c}/\Gamma_\text{e}$
approaches unity for low temperatures, a horizontal hook appears in the plot of
the ratio of Eddington factors in Fig.~\ref{eddtep} due to a radially decreasing
bound-free opacity and a local temperature maximum at the boundary between the
photosphere and the wind. For temperatures lower than $\teff$, the ratio
decreases to one, indicating the diminishing role of bound-free transitions in
low-density supersonic winds.

\begin{figure}
\includegraphics[width=0.5\textwidth]{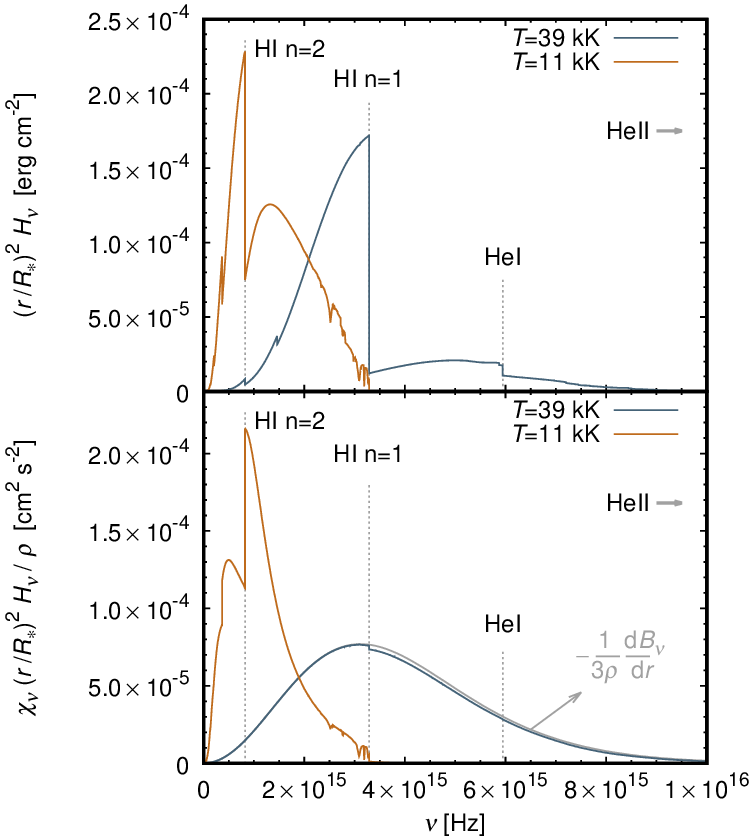}
\caption{Frequency distribution of radiative quantities in the model M1T15
(described in Table~\ref{hvezparlim}), overplotted with ionisation edges (dotted grey lines with arbitrary length). We plot this at two locations in the atmosphere
with different local temperatures (blue and brown lines), roughly
corresponding to the maxima of the relative Eddington parameter in
Fig.~\ref{eddtep}. {\em Top}: Eddington flux scaled relatively to the
stellar radius. {\em Bottom}: Product of the scaled Eddington flux and
opacity, which determines the radiative acceleration. This product is roughly
given by the derivative of the Planck function (grey curve) in
Eq.~\eqref{fraddif} in the optically thick part of the atmosphere (for
$T=39\,\text{kK}$).}
\label{t15311g234y0851}
\end{figure}

The attribution of the Eddington parameter maximum at a temperature of about
$39\,\text{kK}$ to hydrogen ionisation might seem surprising because \ion{He}{ii}
ionisation occurs around this temperature. However, the radiative force is given
by the integral of the product of flux $H_\nu$ and opacity $\chi_\nu$, which in
the diffusion approximation \citep[page 375]{hubenymihalas} gives
\begin{multline}
\label{fraddif}
f_\text{rad}=\frac{4\pi}{c}\int_0^\infty \chi_\nu H_\nu\,\de\nu\approx
\frac{4\pi}{3c}\int_0^\infty \chi_\nu \frac{\de B_\nu}{\de\tau_\nu}\,\de\nu\\=-
\frac{4\pi}{3c}\frac{\de T}{\de r}\int_0^\infty \frac{\de B_\nu}{\de T}\,\de\nu
= - \frac{16\sigma_\text{R}}{3c}T^3\frac{\de T}{\de r},
\end{multline}
where $\sigma_\text{R}$ is the Stefan-Boltzmann constant. Within the diffusion
limit, the radiative flux becomes inversely proportional to the opacity. This
cancels out the opacity dependence of the radiative force, and the radiative
force becomes proportional to the gradient of the Planck function
(Fig.~\ref{t15311g234y0851}, lower plot). In other words, most of the flux is
transferred in regions with low opacity (Fig.~\ref{t15311g234y0851}, upper
plot). As a result, the radiative force does not significantly depend on the
opacity at a given frequency, but rather on the temperature gradient, which is
given by the mean Rosseland opacity. This opacity is a harmonic mean, to which
the lowest opacities contribute most. In turn, the Eddington parameter is
determined by the low opacity around the flux maxima, which is due to hydrogen.
The flux above the \ion{He}{ii} ionisation threshold is low, and therefore, helium
does not significantly contribute to the continuum radiative force around
$39\,\text{kK}$.

Despite the relatively strong effect of the continuum radiative force on the
subsonic region, the contribution of other continuum processes in addition to the light
scattering on free electrons to the wind driving is small. The ratio of the total
continuum to the free-electron radiative force approaches one in the wind, which
means that the direct effect of bound-free and free-free transitions on the wind
mass-loss rates is small for the stars studied here. The same conclusion already
follows from Table~\ref{hvezparlim}, which shows that the models with limited
and unlimited continuum radiative force predict nearly the same wind parameters.
This could change for stars where the wind is driven already from the
photosphere. Moreover, the continuum radiative force might modify the mass-loss
rates indirectly due to increased turbulence \citep{jian,deb2d}.

\section{Discussion}

\subsection{Approaching the Eddington limit}

A fraction of CSPNe show emission-line spectra. In analogy with massive stars,
they are hydrogen deficient and classified as Wolf-Rayet ([WR]) stars \citep{wrwh}. The stellar
winds of massive Wolf-Rayet stars are thought to start their acceleration deep within
the atmosphere, in the optically thick region of the so-called iron opacity bump
\citep[e.g.][]{sanpreh}. These stars have high wind efficiencies, $\eta>1$,
defined as
\begin{equation}
\label{eta}
\eta=\frac{\dot Mv_\infty}{L/c}
\end{equation}
\citep[who used $\varepsilon$ for this quantity]{abbvln}. Most of the stars
we studied have low wind efficiencies, $\eta\lesssim0.1$, and their winds
are therefore optically thin in the continuum \citep{vigratran}.

An estimate of the luminosity at which the wind efficiency reaches unity can be
derived from a fit of the predicted mass-loss rates as a function of the
luminosity,
\begin{equation}
\log\zav{\frac{\dot M}{1\, \msr }}=  -15.45+ 1.97\log\zav{\frac{L}{L_\odot}}.
\end{equation}
The fit was derived for stars with $\teff>25\,\text{kK}$ (above the bistability
jump). When this is combined with Eq.~\eqref{eta}, the limiting luminosity
\mbox{($\eta=1$)} above which the winds are optically thick in the continuum
becomes
\begin{equation}
\log\zav{\frac{L}{L_\odot}}=4.91-1.03\log\zav{\frac{v_\infty}{10^3\,\kms}}.
\end{equation}
When we take the typical CSPN parameters into account \citep{milbe}, it can be
concluded that hydrogen-dominant CSPNe never reach the optically thick limit
and that their winds can be considered as continuum optically thin.  Because
changing the helium mass fraction does not significantly affect the mass-loss
rate \citep[unless close to the Eddington limit due to free electrons,]
[]{vinkolbv,wrtep,esosubwind2}, the appearance of [WR] stars with a continuum
optically thick stellar wind requires a substantial enrichment of the stellar
surface with products of nuclear reactions. This supports the evolutionary
filiation between [WR] CSPNe and stars enriched through a late helium-shell flash
or final thermal pulse \citep{wrwh}.

The origin of the [WR] phenomenon is further supported by the spectral analysis
of planetary nebulae with WR-type central stars \citep{gomgonzolaplan1,
gomgonzolaplan3, gomgonzolaplan2,vkuk}. These stars do not show elevated
luminosities in comparison to hydrogen-rich CSPNe, making the overabundance of
heavier elements the most likely explanation of their strong mass loss (assuming
that [WR] CSPNe have similar masses as hydrogen-rich stars).

\subsection{Connecting stellar and wind models}

The lower boundary of our wind simulations is based on the stellar radius and
effective temperature taken from evolutionary models. This approach works well
for stars with thin atmospheres, where a steep density gradient does not provide
much freedom in the exact location of the connection point between the stellar
and wind model. However, this does not hold for stars with high luminosities,
where the extent of the atmosphere becomes comparable to the stellar radius.

With $\Gamma_\text{e}=0.1-0.3$, the atmospheres are rather far away
from the classical Eddington limit, even for the most luminous stars from our
sample. The bound-free radiative force changes this, however. The generalised Eddington
factor, which includes all continuum opacity, can reach up to $ \Gammac =0.7$
for stars with high luminosities. For these stars, the extension of the
atmosphere becomes comparable to the stellar radius, and the exact location of
the connection point between the stellar and wind models becomes important
\citep{slepovani}. Therefore, CSPN wind models at luminosities higher than those
studied here might require a more elaborate lower boundary condition.

\section{Conclusions}

We used our global wind code METUJE to study the luminosity dependence of the
stellar winds of central stars of planetary nebulae. The code predicts the wind
models just from the stellar parameters and allows for a consistent transition
from the nearly hydrostatic photosphere to the radiatively driven stellar wind.
In particular, the code predicts the wind mass-loss rates and terminal
velocities, which can be tested against observational data, and emergent fluxes,
which are important for the nebular modelling. For our study, we adopted the
stellar parameters from evolutionary calculations of post-AGB stars.

We showed that the mass-loss rate mainly strongly depends on the stellar
luminosity, while the wind terminal velocity increases with the effective
temperature due to the accompanying evolutionary decrease in the stellar radius.
We provided fits for the wind mass-loss rate and the terminal velocity
parameterised in terms of the stellar luminosity and effective temperature. A
comparison with observations showed that although there is large scatter, our
models reproduce the observed mass-loss rates reasonably well. The predicted wind
terminal velocity also compares well with the observed trend of an increase in
the wind terminal velocity with the effective temperature, although the
predicted terminal velocities are slightly lower than their observational
counterparts.

We provided a number of ionising photons for individual models. The comparison
with plane-parallel models showed that winds do not affect the ionisation fluxes
when the ionisation frequency is located below the maximum of the photospheric
flux distribution. However, in the opposite case, the ionising radiation is
strongly modified by the wind. As a result, the  number of \ion{He}{i} and
\ion{H}{i} ionising photons for stars cooler than about
$\teff\approx50\,\text{kK}$ and $\teff\approx40\,\text{kK}$, respectively,
should be calculated using global models. Plane-parallel models never give the
correct \ion{He}{ii} ionising fluxes for the stars considered here.

We focused on stars approaching the Eddington limit. In the models we studied, the
generalised Eddington factor accounting for the contribution of all continuum
transitions can reach up to 0.7 in the opacity maxima at 10 and 39\,kK due to
hydrogen ionisation from $n=2$ and $n=1$ levels, respectively. We showed that
the derived luminosity dependence of the mass-loss rate predicts optically thick
winds with high efficiencies only for stellar luminosities that far exceed typical
CSPN luminosities. This indicates that [WR]-type centrals star originate in
objects with surface layers that were enriched with heavier elements during previous
evolutionary phases.

\section*{Data availability}

Wind and atmosphere models together with mass-loss rate routines are available
at https://zenodo.org/records/23975349.

\begin{acknowledgements}
This work was supported by grant GA \v{C}R 25-15910S. Computational resources
were provided by the e-INFRA CZ project (ID:90254), supported by the Ministry of
Education, Youth and Sports of the Czech Republic. The Astronomical Institute
Ond\v{r}ejov is supported by a project RVO:67985815 of the Academy of Sciences
of the Czech Republic.
\end{acknowledgements}

\bibliographystyle{aa}
\bibliography{papers}

\end{document}